\documentclass[preprint, authoryear]{elsarticle}

\usepackage{natbib}
\usepackage[utf8]{inputenc} 
\usepackage[T1]{fontenc}
\usepackage{lmodern}
\usepackage[fleqn]{mathtools}
\usepackage{booktabs}
\usepackage{rotating}
\usepackage{amssymb}
\usepackage{amsmath}
\usepackage{graphicx}
\usepackage{caption}
\usepackage{epstopdf}
\usepackage{stackrel}
\usepackage{soul}
\usepackage{color}
\usepackage{float}
\usepackage[official]{eurosym}
\usepackage{tabularx}
\usepackage{multirow}
\usepackage{makecell}
\usepackage{lscape}
\usepackage{afterpage}
\usepackage{url}
\usepackage{mathrsfs}
\usepackage{chngcntr}
\usepackage{caption}
\usepackage{subcaption}
\usepackage{wasysym}
\usepackage{xfrac}
\usepackage[table]{xcolor}

\usepackage{ulem}
\usepackage{todonotes}

\usepackage{comment}

\newcommand*\diff{\mathop{}\!\mathrm{d}}

\newdefinition{rmk}{Remark}
\newproof{pf}{Proof}
\newproof{pot}{Proof of Theorem \ref{thm2}}

\makeatletter
\def\ps@pprintTitle{%
	\let\@oddhead\@empty
	\let\@evenhead\@empty
	\def\@oddfoot{\centerline{\thepage}}%
	\let\@evenfoot\@oddfoot}
\makeatother

\begin{document}
\pagenumbering{gobble}
\title{A closer look at the chemical potential of an  \\ ideal agent system \\[12pt]
{\small(Version: \today)}}

\author[hhu]{Christoph J.\ Börner}
\ead{Christoph.Boerner@hhu.de}	
\author[hhu]{Ingo Hoffmann\corref{cor1}}
\ead{Ingo.Hoffmann@hhu.de}	
\author[hhu]{John H.\ Stiebel}
\ead{John.Stiebel@hhu.de}

\cortext[cor1]{Corresponding author. Tel.: +49 211 81-15258; Fax.: +49 211 81-15316}

\address[hhu]{Heinrich Heine University D\"usseldorf, Faculty of Business Administration \\ and Economics, Financial Services, 40225 D\"usseldorf, Germany,
\\ ROR: https://ror.org/024z2rq82}	

\begin{abstract}
Models for spin systems known from statistical physics are used in econometrics in the form of agent-based models.	
Econophysics research in econometrics is increasingly developing general market models that describe exchange phenomena and use the chemical potential $\mu$ known from physics in the context of particle number changes. 
In statistical physics, equations of state are known for the chemical potential, which take into account the respective model framework and the corresponding state variables.
A simple transfer of these equations of state to problems in econophysics appears difficult. To the best of our knowledge, the equation of state for the chemical potential is currently missing even for the simplest conceivable model of an ideal agent system. In this paper, this research gap is closed and the equation of state for the chemical potential is derived from the econophysical model assumptions of the ideal agent system. An interpretation of the equation of state leads to fundamental relationships that could also have been guessed, but are shown here by the theory.
\end{abstract}

\begin{keyword}
	Agent System
	\sep Chemical Potential
	\sep Econophysics
	\sep Entropy
	\sep Gibbs Free Energy \\[6pt] 
	
	
	\noindent \textit{ORCID IDs:} 
	0000-0001-5722-3086 (Christoph J.~B\"orner), 
	0000-0001-7575-5537 (Ingo Hoffmann), 
	0000-0003-0088-2456 (John H.~Stiebel),    	\\[6pt] 
	
\end{keyword}
\maketitle
\newpage

\pagenumbering{arabic}

\section{Introduction} \label{Introduction}
The chemical potential $\mu$ is a very well known state variable in statistical physics \citep{Greiner.1995} and is considered in cases where a many-particle system is not only in exchange with a heat bath but also with a particle reservoir \citep{Isihara.1971, Landau.1980}.
Such systems are described in the grand canonical ensemble \citep{Kardar.2007},
where the associated partition function is the grand canonical partition function \citep{Isihara.1971, Kardar.2007} and the associated thermodynamic potential is the free enthalpy $G$ \citep{Fliebach.2018}.
The chemical potential has a particularly simple relationship to the free enthalpy: $G = N \mu$ if $N$ denotes the number of particles and the other state variables of the system are constant \citep{Fliebach.2018}. In some literatures, $G$ is also referred to as Gibbs free enthalpy \citep{Greiner.1995} or Gibbs free energy \citep{Isihara.1971, Kardar.2007}.
It is summarized that the chemical potential corresponds to the energy required to add another particle to a system if all other state variables are constant \citep{Greiner.1995, Shegelski.2004, Fliebach.2018}.

Many methods of statistical physics have found their way into econometrics and sociology, where they lead to models, e.g., for a large number of investors or decision-makers, which are generally summarized under the generic term agents. The corresponding field of research is now generally referred to as econophysics \citep{Chakraborti.2011, Chakraborti.2011b}. 
Starting with the work of \citet{Weidlich.1971} and \citet{Galam.1982}, the research field of econophysics utilizing agent systems has steadily developed and branched out; see, e.g.\ \citet{Kaizoji.2000, Kozuki.2003, Oh.2007, Kleinert.2007, Kozaki.2008, Krause.2012, Bouchaud.2013, Sornette.2014, Crescimanna.2016}.

A starting point in econophysics is the assumed analogy between agents and spins known in physics in a lattice arrangement. The analogously used model class is recruited from the Ising models \citep{Ising.1925} for spins; see, e.g.\ \cite{Bouchaud.2013} and  \cite{Sornette.2014} for a review. Extensions are also considered in which the number of agents is not constant, and an overall modeling with exchange processes is discussed, see, e.g.\ \citet{Saslow.1999, Sousa.2006}. 
In some of these overall models, the chemical potential $\mu$ is introduced as a constant parameter and used to model the exchange processes, see, e.g.\ \citet{Staliunas.2003, Schmidhuber.2022}.
It is known from statistical physics that the chemical potential -- depending on the state variables of the system -- can be described by an equation of state, for example the concentration dependence of $\mu$ in chemical reactions of gases \citep[Eq.\ 3.36]{Greiner.1995}.

If approaches based specifically on Ising models are considered in econophysics, then one model component describes the interaction of the agent with a message field and one component describes the interaction of the agents with each other. If the interaction of the agents with each other tends towards zero, i.e.\ the agents act in isolation, then the corresponding models describe a so-called ideal agent system \citep{Bouchaud.2013, Sornette.2014, Boerner.2023, Boerner.2023c, Boerner.2023b}.

Despite an intensive search, no reference to an equation of state for the chemical potential in ideal agent systems could be found in the literature of econophysics. This article aims to close this research gap. The equation of state for the chemical potential is derived for the simplest conceivable ideal agent system with two states.
General implications are deduced from this equation of state, which could also have been guessed, but are derived here from the theory.

This study is structured as follows.
The next section provides a brief overview of the specific literature. Related research strands in econophysics are highlighted and a delimitation of the present contribution is given. In section \ref{AgentSystems}, the equation of state for the chemical potential for an ideal agent system is derived and in section \ref{Application} a capital market application is shown. Section \ref{ImplicationsLimits} is dedicated to the limitations of the application and describes general conclusions from the equation of state before a conclusion is made in section \ref{Conclusion}.

\section{Literature and Delimitation} \label{Literature}
In addition to approaches based on spin lattices that build a bridge to econometrics through analogy, see, e.g.\ \citet{Kaizoji.2000, Kozuki.2003, Oh.2007, Kleinert.2007, Kozaki.2008, Krause.2012, Bouchaud.2013, Sornette.2014, Crescimanna.2016}, there are a variety of contributions that deal with the transfer of models from statistical physics for gases to econometrics, see, e.g.\ \citet{Saslow.1999, Sousa.2006, Bicci.2016, Abel.2020, Braten.2021} and the many other literature sources listed there.
In the latter publications, the chemical potential is introduced in different contexts and used in different econometric models. The interpretation of the meaning of the chemical potential is, as far as can be seen, dependent on the context. An equation of state for the chemical potential is not considered there; only an approach to calculate changes in the chemical potential $\Delta \mu$ when the density $\rho$ of the agents changes, i.e.\ $\rho_1 \rightarrow \rho_2$, can be found in \citet[Eq.\ 21]{Braten.2021}.

An extension of the gas models in statistical physics is the consideration of gases made up of particles with a half-integer or an integer spin value. For which the Fermi-Dirac statistics or the Bose-Einstein statistics apply and which take into account effects such as the Pauli exclusion principle or the Bose-Einstein condensation, see, e.g.\ \citet{Isihara.1971, Landau.1980, Huang.1987, Greiner.1995, Kardar.2007}.
The statistical physics models for these special gases are also discussed in econometric applications, see, e.g.\ \citet{Chatterjee.2003, Staliunas.2003, Kuerten.2012, Rashkovskiy.2019} and the literature cited therein. 
As far as can be seen, no equations of state for the chemical potential are considered in the econophysical analysis of these gas systems. Whereas in physics they are derived in special applications, see, e.g.\ \citet{Wilson.1931, Shegelski.2004, Sevilla.2011}.

A simple transfer of the equations of state for the chemical potential in physics into econometrics by analogy is likely to be difficult, 
since the chemical potential depends on other state variables (for example: volumes, energy levels, effective masses) that -- yet -- have no meaning in econometrics.
Presumably the only viable way in the corresponding applications in econometrics is to derive equations of state for the chemical potential from the context and to use concepts from physics.

For agent systems based on spin lattice approaches, exchange phenomena in econometrics are also considered and modeled with a chemical potential, see, e.g.\ \citet{Lemoy.2011, Murota.2014, Schmidhuber.2022}. However, as far as can be seen, no equation of state is given for the chemical potential.

The basic model here is an Ising model \citep{Ising.1925} and, in principle, all state variables and model parameters considered in econometrics are measurable variables, albeit with greater effort, see, e.g.\ \citet{Boerner.2023, Boerner.2023c, Boerner.2023b}.
This line of research is the focus of the present contribution and will be further developed and the equation of state for the chemical potential for the simplest example of an agent system depending on its state variables will be specifically derived. Concepts from physics are used for derivation, see, e.g.\ \citet{Greiner.1995, Fliebach.2018}.

\section{Two-State Ideal Agent System}\label{AgentSystems}
\subsection{Model}\label{Model}
An agent system based on the Ising model \citep{Ising.1925} without interaction between the agents is described below, see, e.g.\ \cite{Bouchaud.2013, Sornette.2014, Boerner.2023c}. Following \citet{Boerner.2023b} we briefly summarize the model. 

A number $N$ of participants will be considered as an example of a system of agents. A news situation $-1\leq {\cal B} \leq +1$ is given and influences the agents. For the sake of simplicity, it is assumed that each agent only acts based on the news situation ${\cal B}$.
In capital market applications, for example, the news situation triggers the buy or sale of a share. Agreements or alliances between agents are excluded, so each agent acts in isolation and is uninfluenced by other agents. These systems can be referred to as systems without interactions or ideal agent systems; see, e.g.\ \citet{Bouchaud.2013, Boerner.2023c}.

Without loss of generality, the following will be limited to positive news situations, i.e.\ the strength $|{\cal B}| = B$ for the news situation is assumed.
Agents can act in "conformity" to the news, i.e.\ investors buy if the news is positive, or in "non-conformity", i.e.\ investors sell if the news is positive. The state of an agent $i = 1, \ldots, N$ is described with $s_i = +1$ (conform) or $s_i = -1$ (non-conform).
If a dynamic equilibrium is established, then there is a number $N_{+}$ of "conform" and a number $N_{-}$ of "non-conform" agents, with $N = N_{+} + N_{-}$. If an inversion of the agent system is ruled out, i.e.\ the overall system acts in accordance with the positive news, the difference $ N_{+} - N_{-}$ is positive and corresponds to a surplus of "conform" agents.

In econometric applications, the normalized variable ${\bar N}_{\rm pot} = \frac{1}{N} (N_{+} - N_{-}) = \frac{1}{N} \sum_i s_i $ can be referred to as trade potential ({\it phys.}: $\propto$ magnetization) and can be used, e.g., for one-step ahead forecast models \citep{Vikram.2011, Boerner.2023c}.

\subsection{Energie, Utility, Entropy and Factor $k$}\label{Utility}
As summarized in \citet{Boerner.2023c} analogy considerations establish a connection between energy $E$ known from physics and utility $U$ in econometrics. While phyiscal systems minimize their energy, utility maximization is assumed in econometrics. Hence, as a first starting point for further calculations, the relation $E = -U$ is often found in the econophyics literature \citep{Marsili.1999, Sornette.2014, Boerner.2023c} and provides a theoretical framework for transferring concepts from statistical physics to econometrics.

Nevertheless, models based on analogies are limited. Their application to systems that have more complex dynamics, such as segregations \citep{Schelling.1971, Schelling.1978}, must be questioned, especially if individual utility are also taken into account \citep{Grauwin.2009,Lemoy.2011, Bouchaud.2013}. These extensions are not taken into account in this contribution and we assume the relation $U=-E$ as the starting point.
\subsubsection*{Utility}
Thus, if only the interaction of the agents with the message environment is considered, following \citet{Boerner.2023c}, the utility of a specific configuration of the ideal agent system is
\begin{flalign} \label{UtilityEq}
	U =  \mu_{\rm B} B (N_{+} - N_{-}),
\end{flalign}
where $\mu_{\rm B}$ ({\it phys.}: magnetic moment or Bohr magneton) denotes the parameter that describes the increase in utility per "conform" agent in a message environment $B$. In order to avoid confusion with the chemical potential, the parameter $\mu_{\rm B}$ with index ${\rm B}$ is used as a reference to the news field $B$. \citet{Boerner.2023c} show one way to interpret the utility $U$ and determine the parameter $\mu_{\rm B}$ in a capital market example. Analogous to physics, the quantity $M = \mu_{\rm B} (N_{+} - N_{-})$ can be referred to as magnetization in econophysics. With the magnetization $U = M B$ follows. Application-related designations for the magnetization $M$ can be found in \citet{Michard.2005, Vikram.2011, Bouchaud.2013, Zhang.2015b}.

\subsubsection*{Entropy}
If the microcanonical partition function $\Omega = \frac{N!}{N_{+}!N_{-}!}$, see, e.g.\ \citet{Boerner.2023c}, and equation (\ref{UtilityEq}) is used, the entropy $S = S(U, B, N) = -k\ln\Omega$ can be calculated for the ideal agent system as a function of the utility $U$, the message environment $B$ and the number $N$ of agents:
\begin{flalign} \label{EntropyEq}
	S =  -kN \left( p_{+}\ln p_{+} + p_{-}\ln p_{-}\right) \quad {\rm with}\;  p_{\pm} = \left(\frac{1}{2} \pm \frac{U}{2\mu_{\rm B} B N}\right).
\end{flalign}
Except for the factor $k$ ({\it phys.}: Boltzmann constant), this expression is found identically in \citet[Eq.\ 6.24]{Fliebach.2018}.
Note, $M_0 =\mu_{\rm B} N$ describes the maximum magnetization of the agent system and $U_0 = \mu_{\rm B} B N = M_0 B $ then describes the maximum utility, i.e.\ all agents act "conform" to the news field $B$.

\subsubsection*{Factor $k$}

In physics, the Boltzmann constant $k$ transforms the state variable temperature $T$ from Kelvin to Joules and is therefore a proportionality constant so that balance equations can be calculated in equal units of energy \citep{Greiner.1995}. In equation (\ref{EntropyEq}) the constant $k$ is not a model parameter but defines also the unit of measurement of the utility \citep{Boerner.2023c}. The question is, how can the factor $k$ be determined in econophysics?

Equation (\ref{EntropyEq}) is very similar to the Shannon entropy ${\cal S}$ known from information theory \citep{Shannon.1948}. The entropy from equation (\ref{EntropyEq}) can be represented with the factor $\sfrac{1}{N \ln 2}$ per agent and in units of bit \citep{Nadaletal.1998}. With the occupation probabilities $p_{\pm}$ from equation (2) it follows:
\begin{flalign} \label{EntropyEq_2}
	{\cal S} =  -k \left( p_{+}\log_2 p_{+} + p_{-}\log_2 p_{-}\right).
\end{flalign}
Complete uncertainty about the state ("conform" or "non-conform") of an agent exists when the utility $U$ in equation (\ref{EntropyEq}) approaches zero ({\it phys.}: high temperature limit). The occupation probabilities $p_{\pm}$ are then equal and have the value $p_{\pm} = \sfrac{1}{2}$. This means that the bracket in equation (\ref{EntropyEq_2}) then becomes $-1$ and the Shannon Entropy ${\cal S}$ per agent takes on the value $k$.

Shannon introduced $k$ as a positive constant and assigned it the property of a unit of measurement of information \citep[p.\ 11]{Shannon.1948}.
Thus, one way to determine the value of $k$ in practice is to obtain an answer to the following question: 
How much is the information about the exact state of an agent worth to an observer (researcher, market analyst, investor, reporter, etc.; see, e.g.\ \citet{Boerner.2023c}) and in which monetary units is the value measured if the agent system is in a state of maximum uncertainty?

In \citet{Boerner.2023c}, the value is set as an example at $k = 1$ USD and is adopted here.

\subsection{Caloric and thermal equation of state}\label{CalThermEqOfStat}
Suitable derivations of the entropy equation (\ref{EntropyEq}) lead to equations for the ideal agent system that correspond to the caloric and thermal equations of state in statistical physics.
\subsubsection*{Caloric equation of state}
With the derivation 
\begin{flalign} \label{ÖkoTemp}
	\frac{1}{T} :=  - \left. \frac{\partial S}{\partial U}\right|_{{B, N} = {\rm const.}}
\end{flalign}
introduced by \citet{Marsili.1999} into econophysics, a state variable $T$ is defined for the agent system, which corresponds to the temperature in statistical physics; see also \citet{Boerner.2023b} specifically for the ideal agent system.
The evaluation of the derivative in equation (\ref{ÖkoTemp}) leads to the following equation:
\begin{flalign} \label{CaloricEq_1}
	\frac{\mu_{\rm B} B}{kT} =  \tanh^{-1}\left(\frac{U}{\mu_{\rm B} B N}\right).
\end{flalign}
For the quotient $\frac{U}{\mu_{\rm B} B N}$, the order of magnitude $10^{-2}$ and smaller is found in \citet{Boerner.2023c}, i.e.\ the utility $U$ is significantly smaller in practice than the theoretical, maximum utility $U_0$. Thus, the inverse hyperbolic tangent  function for small arguments can be developed in a Taylor series, $\tanh^{-1}(x) \approx x$ for $x\approx 0$, and the caloric equation of state is calculated:
\begin{flalign} \label{CaloricEq_2}
	U(T, B, N) =  \frac{N(\mu_{\rm B} B)^2}{kT}.
\end{flalign}
\subsubsection*{Thermal equation of state}
Carrying out the derivative
\begin{flalign} \label{ÖkoMagTemp}
	\frac{M}{T} :=  \left. \frac{\partial S}{\partial B}\right|_{{U, N} = {\rm const.}}
\end{flalign}
leads to the equation:
\begin{flalign} \label{ThermalEq_1}
	\frac{M}{kT} =  \frac{U}{\mu_{\rm B} B^2}\tanh^{-1}\left(\frac{U}{\mu_{\rm B} B N}\right).
\end{flalign}
If the same approximation as before is used for the inverse hyperbolic tangent function and the caloric equation of state, equation (\ref{CaloricEq_2}), is used, the thermal equation of state is calculated:
\begin{flalign} \label{ThermalEq_2}
	M(T, B, N) =  \frac{N \mu_{\rm B}^2}{k} \frac{B}{T}.
\end{flalign}
Note that a further partial derivative with respect to $B$ leads to a quantity known in physics as magnetic susceptibility. The right side of the equation then reflects the well-known Curier law, see, e.g.\ \citet[Eq.\ 26.15]{Fliebach.2018}. The quantity equivalent to the Curier constant is abbreviated as $\gamma = \sfrac{N \mu_{\rm B}^2}{k}$ to simplify the following calculations.

\subsection{Polarization entropy}\label{PolEntropy}
 As an intermediate variable for determining the equation of state for the chemical potential, the entropy $S(U, B, N)$ is calculated as a function of the state variables $T$ and $M$ with constant $N$. Thus, $S = S(T, M)$ is sought in the following.

With a constant number of agents, i.e.\ $\diff N = 0$, follows with equations (\ref{ÖkoTemp}) and (\ref{ÖkoMagTemp}) for the total differential of entropy $S = S(U, B)$:
\begin{flalign} \label{TotalDiffS_Eq}
	\diff S =  -\frac{1}{T}\diff U + \frac{M}{T} \diff B.
\end{flalign}
The substitution of the caloric and thermal equations of state, equations (\ref{CaloricEq_2}) and (\ref{ThermalEq_2}), lead to:
\begin{flalign} \label{TotalDiffS_Eq2}
	\diff S =  -\frac{1}{\gamma} M \diff M.
\end{flalign}
Starting from an arbitrary reference state $T_a$, $M_a$ with entropy $S_a$, this equation is integrated:
\begin{flalign} \label{PolEntropy_Eq1}
	S(T,M) = \underbrace{S_a(T_a,M_a) + \frac{1}{2\gamma} M_a^2}_{\rm reference\; state} - \frac{1}{2\gamma} M^2.
\end{flalign}
It can be seen in equation (\ref{PolEntropy_Eq1}) that only a direct dependence on the magnetization $M$ of the ideal agent system determines the entropy. This result is consistent with the results of \citet{Redfield.1969}. For a system of independent nuclear spins, heuristic techniques and empirical conciderations led him to the same equation for the polarization entropy \citep[Eq.\ 6]{Redfield.1969}, but without considering a reference state.
Although the arbitrarily chosen reference state is usually of no importance in physics, since only entropy differences between two states are considered and the reference state only determines the absolute entropy scale \citep{Greiner.1995}, we will specify the reference state later.	

\subsection{Gibbs free energy and Gibbs-Duhem-Relation} \label{GibbsFreeEnergy}
Rearranging equation (\ref{TotalDiffS_Eq}) leads to the total differential for the utility $U$. If a change in the number of agents $\diff N$ is considered, a positive additive utility should occur if additional agents are to be taken into account. This increase in utility is evaluated with the still unknown chemical potential $\mu$ per agent. The total differential of $U$ is expanded accordingly:
\begin{flalign} \label{TotalDiffU_Eq}
	\diff U =  -T \diff S + M \diff B + \mu \diff N.
\end{flalign}
If two Legendre transformations are carried out ($S \rightarrow T$ and $B \rightarrow M$), this leads to the total differential of the Gibbs free energy:
\begin{flalign} \label{TotalDiffG_Eq}
	\diff G =  +S \diff T - B \diff M  + \mu \diff N.
\end{flalign}
On the other hand, the following simple relationship applies to the Gibbs free energy \citep{Fliebach.2018}: $G(T, M, N) = N\mu(T,M)$. Thus the total differential of $G$ becomes:
\begin{flalign} \label{TotalDiffG_Eq2}
	\diff G =  +N \diff \mu  + \mu \diff N.
\end{flalign}
Equating equations (\ref{TotalDiffG_Eq}) and (\ref{TotalDiffG_Eq2}) leads to an equation analogous to the Gibbs-Duhem relation known from physics \citep{Greiner.1995}:
\begin{flalign} \label{TotalDiffMU_Eq1}
	\diff \mu =  +\frac{S}{N}\diff T - \frac{B}{N} \diff M.
\end{flalign}

\subsection{Equation of state of the chemical potential}\label{ChemPot}
In equation (\ref{TotalDiffMU_Eq1}), equation (\ref{PolEntropy_Eq1}) is first substituted and then the thermal equation of state, equation (\ref{ThermalEq_2}), in the form $B = \sfrac{1}{\gamma} MT$. This leads to:
\begin{flalign} \label{TotalDiffMU_Eq2}
	\diff \mu =  \frac{1}{N}
	 \left( S_a(T_a,M_a) + \frac{1}{2\gamma} M_a^2 - \frac{1}{2\gamma} M^2 \right)\diff T 
				-\frac{1}{N}
	 \frac{1}{\gamma} MT 	\diff M	.
\end{flalign}
Starting again from an arbitrary reference state $T_a, M_a$ with chemical potential $\mu_a$, equation (\ref{TotalDiffMU_Eq2}) is integrated:
\begin{flalign} \label{ChemPot_Eq1}
	\mu(T,M) = \mu_a(T_a,M_a) + \frac{1}{N}S_a(T_a,M_a)(T - T_a) + \frac{1}{2\gamma N} (M_a^2 - M^2)T.
\end{flalign}

\subsubsection*{Determination of the reference state}
Equation (\ref{TotalDiffU_Eq}) is the total differential for the utility, so that an equation analogous to the Euler equation known from physics \citep{Greiner.1995} can be noted for the ideal agent system:
\begin{flalign} \label{EulerU_Eq}
	U = -TS + MB + \mu N.
\end{flalign}
This equation suggests that for an observer \citep{Boerner.2023c} of the ideal agent system, the utility increases, with $\mu > 0$, the more agents $N$ are involved and when the agents are predominantly "conform", i.e.\ $M>0$, with the message field $B$. The utility decreases the higher the temperature $T$ and the higher the entropy $S$ of the ideal agent system. Both are measures of uncertainty, so the utility decreases as uncertainty increases.

With the caloric and thermal equations of state, equations (\ref{CaloricEq_2}) and (\ref{ThermalEq_2}), $U = MB$ follows for an ideal agent system. Thus
\begin{flalign} \label{EulerU_Eq2}
	TS = \mu N.
\end{flalign}

Equations (\ref{PolEntropy_Eq1}) and (\ref{ChemPot_Eq1}) are substituted into this equation and the condition for the reference states follows:
\begin{flalign} \label{RefStateCond_Eq}
	T_a S_a(T_a, M_a) = N \mu_a(T_a, M_a).
\end{flalign}
Since this equation no longer depends on $T$ and $M$, Euler's equation (\ref{EulerU_Eq}) for an ideal agent system is always valid, if equation (\ref{RefStateCond_Eq}) holds for the reference states. Here, $T_a$ and $M_a$ for the reference states are
completely arbitrary and freely selectable. 
If $T_a = 0$ is chosen, a maximum magnetization $M_a = M_0 = \mu_{\rm B} N$ of the ideal agent system \citep{Bouchaud.2013, Sornette.2014, Boerner.2023c} follows for a non-zero message field, i.e.\ $B \neq 0$. This means that the entropy becomes zero and the following applies: 
$S_a(T_a, M_a) = S_a(0, M_0) = 0$. With equation (\ref{RefStateCond_Eq}) $\mu_a(T_a, M_a) = 0$ follows immediately.

\subsubsection*{Equation of state}
With the definition of the reference states it follows from equation (\ref{ChemPot_Eq1}) the equation of state for the chemical potential:
\begin{flalign} \label{ChemPot_Eq2} \nonumber
	\mu(T,M) & = \frac{1}{2\gamma N} (M_0^2 - M^2) T  
			\qquad {\rm with}\;     M_0 = \mu_{\rm B} N 
			\;    {\rm and} \;  \gamma = \frac{N \mu_{\rm B}^2}{k}\\
			 & = \frac{1}{2} kT \left( 1 - \frac{M^2}{M_0^2}\right).
\end{flalign}

\section{Application}\label{Application}
A capital market example shows a possible application. Here the ideal agent system consists of investors who buy or sell a stock.
From \citet[Eq.\ 9]{Boerner.2023c} follows:
\begin{flalign} \label{CapAppl_Eq_1}
	\frac{M^2}{M_0^2} 
		& = \frac{\mu_{\rm B}^2 (N_{+} - N_{-})^2 }{\mu_{\rm B}^2 N^2} 
		 = {\bar N}_{\rm pot}^2.
\end{flalign}
With the trade potential ${\bar N}_{\rm pot}$ as the balance of buyers and sellers. 
\citet{Boerner.2023c} find empirical values of the order of $10^{-2}$ and smaller for the trade potential, i.e.\ ${\bar N}_{\rm pot}^2 \approx 0$,
and equation (\ref{ChemPot_Eq2}) becomes:
\begin{flalign} \label{ChemPot_Eq3} 
	\mu(T) & \approx \frac{1}{2} kT.
\end{flalign}
In \citet[Eq.\ 9]{Boerner.2023b} the connection between temperature $T$ and volatility $\sigma$ of the stock is given for a two-state ideal agent system consisting of $N$ investors.
The functional connection given in  \citet{Boerner.2023b} describes a monotonically increasing function depending on $N$ and $\sigma$. In the vicinity of a selected operating point, a proportional relationship $T\propto \sigma$ can be established using a Taylor expansion.
This means that the chemical potential is proportional to volatility: $\mu \propto \sigma$.
The result means that for a capital market observer (researcher, market analyst, investor, reporter, etc.; see, e.g.\ \citet{Boerner.2023c}) the utility from an additional investor is greater when the volatility increases.

This simple connection could have been guessed quite easily, but here it follows from theory.

\section{Limits and Implications}\label{ImplicationsLimits}
\subsubsection*{Limits}
Strictly speaking, the equation of state for the chemical potential derived here, equation (\ref{ChemPot_Eq2}), applies exclusively to systems in which the agents are isolated and do not interact. Likewise, the chemical potential of systems of agents that contain more complex dynamics due to additional, individual utility contributions may not be correctly described \citep{Grauwin.2009, Lemoy.2011, Bouchaud.2013}. The models considered here apply at one point in time and, analogous to physics, assume a thermodynamic equilibrium. Time courses are not taken into account, so that, for example, memory effects, changes in individual preferences or developments in habitus characteristics are not captured in the models \citep{Bouchaud.2013, Boerner.2023c}. The latter require non-equilibrium thermodynamics approaches, see, e.g.\ \citet{Isihara.1971}, which may lead to modifications of the equation of state for the chemical potential.

In addition to the limits due to the model simplicity, validity limits based on the approximations must be taken into account. In section \ref{CalThermEqOfStat} the $\tanh^{-1}(x)$ for $x\approx 0$ was developed in a Taylor series up to the first order. 
The prerequisite was that the utility is significantly smaller than the theoretical maximum utility, so that the next order of the Taylor series, i.e.\ $\sfrac{x^3}{3}$, was negligible. This is likely to be the case for many applications in practice, but in special applications where this is no longer the case, attention must be paid to that point and a case-by-case analysis should be carried out.

Another important limitation to the applicability of equation (\ref{ChemPot_Eq2}) arises from the fact that when deriving the entropy, equation (\ref{PolEntropy_Eq1}), the full dependence on the number of agents $N$ was not taken into account. To do this, the term $\sfrac{\mu}{T}\diff N$ would have had to be added to the differential in equation (\ref{TotalDiffS_Eq}). The chemical potential $\mu(T, M, N)$ would therefore have to have already been known in order to be able to carry out the integration. In addition, the constant $\gamma$ is not a constant with respect to the number of agents $N$ and increases as the number of agents increases. If $\diff N$ is approximated by $\Delta N$ in practical applications, it must be checked whether $\sfrac{\Delta N}{N}$ is approximately zero in order to keep the repercussions of both causes mentioned to a minimum. If very large systems with many agents are considered and only small increases in agents are observed, then this condition is met.

\subsubsection*{Implications}
Equation (\ref{ChemPot_Eq2}) offers the possibility in many (overall) models based on spin lattices, to no longer only consider the chemical potential as a constant, predetermined parameter \citep{Murota.2014}, but to include it in the model equations depending on the state of the agent system. For weak couplings between agents, this should be an approximation that becomes better in the limit case of vanishing coupling.

If the ideal agent system considered in detail here is understood and applied as a basic model and first approach for decision situations of the general yes/no type \citep{Weidlich.1971, Galam.1982}, then the equation of state of the chemical potential describes the increase in utility for an observer \citep{Boerner.2023c} when an additional agent joins the decision situation.

This allows the Gibbs free energy $G$ to be interpreted in the context of econophysics. It describes the proportion of utility that can theoretically be extracted from the ideal agent system with $N$ agents if only the migration of agents is considered.

In physics, the chemical potential is one of the intensive state variables and does not depend on the number of particles \citep{Greiner.1995}. Equation (\ref{ChemPot_Eq2}) correctly reflects this in econophysics and is independent of the number of agents in a decision situation. Thus, the increase in utility per agent, i.e.\ the value of the chemical potential, does not depend on the number of agents in the decision situation, but rather on the state in which the agent system is. The state variables here are temperature, as a measure of uncertainty, see also the connection to volatility in section \ref{Application}, and the average magnetization, normalized to 1, as a measure of the prevailing opinion.

The dependence on the magnetization at constant temperature is low due to the square term. Changes in already small magnetizations hardly change the value of the almost maximum chemical potential. However, the increase in utility per additional agent decreases dramatically the closer the prevailing opinion is towards 1. 

If the magnetization remains constant, the state variable temperature has the greatest influence on the chemical potential. With increasing uncertainty in the agent system, an additional agent means a greater increase in utility. Conversely, if the uncertainty in the system tends to zero, the increase in utility from an additional agent is small and, proportional to the temperature, tends to zero in the limit case.

\section{Conclusion}\label{Conclusion}
The highly simplified example of a two-state ideal agent system was
studied in detail, and the equation of state for the chemical potential was derived. It turns out that the chemical potential in the ideal system under consideration depends on the state variables temperature and magnetization. The first variable is a measure of the prevailing uncertainty in the system and the second variable is a measure of the extent to which, on average, the agents have adopted a common position that conforms to the news environment. 
The limitations resulting from possible false conclusions through analogies and strong coupling between agents were also discussed and identified as limits to the application of the equation of state.
Nevertheless, the equation of state can serve as a starting point to obtain initial indications for an overall model in which the chemical potential is not a constant quantity but depends on the state variables of the system.

\bibliography{../../020_Literatur/005_CitaviBibTexFile/CitaviHerding}
\end{document}